\documentclass[12pt]{iopart}

\usepackage{microtype}  

\usepackage{graphicx}
\usepackage{xspace}
\usepackage{float}

\usepackage{hyperref}

\newcommand{\beq}{\begin{equation}}
\newcommand{\eeq}{\end{equation}}

\newcommand{\beqn}{\begin{eqnarray}}
\newcommand{\eeqn}{\end{eqnarray}}

\newcommand{\ahfd}{\texttt{AHFinderDirect}\xspace}

\newcommand{\BaikalVacuum}{{\texttt{BaikalVacuum}}\xspace}
\newcommand{\bhahaha}{{\texttt{BHaHAHA}}\xspace}
\newcommand{\bah}{{\texttt{BHaHAHA}}\xspace}
\newcommand{\bhah}{{\texttt{BlackHoles@Home}}\xspace}

\newcommand{\Carpet}{\texttt{Carpet}\xspace}
\newcommand{\dendrogr}{\texttt{Dendro-GR}\xspace}
\newcommand{\einsteintoolkit}{\texttt{Einstein Toolkit}\xspace}
\newcommand{\et}{\texttt{ET}\xspace}
\newcommand{\grchombo}{\texttt{GRChombo}\xspace}

\newcommand{\nell}{{\texttt{NRPyElliptic}}\xspace}
\newcommand{\nrpy}{{\texttt{NRPy}}\xspace}
\newcommand{\openmp}{{\texttt{OpenMP}}\xspace}

\newcommand{\spectre}{\texttt{SpECTRE}\xspace}
\newcommand{\SymPy}{\texttt{SymPy}\xspace}

\newcommand{\eqref}[2][]{Eq#1.~(\ref{#2})}

\usepackage{xcolor}

\makeatletter
\newcommand{\mainmatter}{%
  \setcounter{footnote}{0}%
  \patchcmd{\@makefntext}{\fnsymbol}{\arabic}{}{}%
  \patchcmd{\@thefnmark}{\fnsymbol}{\arabic}{}{}%
  \def\@makefnmark{\textsuperscript{\arabic{footnote}}}%
}
\makeatother

\begin{document}
\date{\today}
\title{\bhahaha: A Fast, Robust Apparent Horizon Finder Library for Numerical Relativity}

\author{Zachariah B.~Etienne}
\address{Department of Physics, University of Idaho, Moscow, ID 83843, USA}
\address{Department of Physics and Astronomy, West Virginia University, Morgantown, WV 26506, USA}
\address{Center for Gravitational Waves and Cosmology, West Virginia University, Chestnut Ridge Research Building, Morgantown, WV 26505, USA}
\ead{zetienne@uidaho.edu}

\author{Thiago Assump\c{c}\~{a}o}
\address{Center for Gravitation, Cosmology and Astrophysics, Department of Physics, University of Wisconsin-Milwaukee, Milwaukee, WI 53211, USA}

\author{Leonardo Rosa Werneck}
\address{Department of Physics, University of Idaho, Moscow, ID 83843, USA}

\author{Samuel D.~Tootle}
\address{Department of Physics, University of Idaho, Moscow, ID 83843, USA}

\begin{abstract}
    Apparent horizon (AH) finders are essential for characterizing black holes and excising their interiors in numerical relativity (NR) simulations. However, open-source AH finders to date are tightly coupled to individual NR codes. We introduce \bah, the \bhah Apparent Horizon Algorithm, the first open-source, infrastructure-agnostic library for AH finding in NR. \bah implements the first-ever hyperbolic flow-based approach, recasting the elliptic partial differential equation for a marginally outer trapped surface as a damped nonlinear wave equation. To enhance performance, \bah incorporates a multigrid-inspired refinement strategy, an over-relaxation technique, and \openmp parallelization. When compared to a na\"ive hyperbolic relaxation implementation, these enhancements result in 64x speedups for difficult common-horizon finds on a single spacetime slice, enabling \bah to achieve runtimes within 10\% of the widely used (single-core) \ahfd and outperform it on multiple cores. For dynamic horizon tracking with typical core counts on a high-performance-computing cluster, \bah is approximately 2.1 times faster than \ahfd at accuracies limited by interpolation of metric data from the host NR code. Implemented and tested in both the \einsteintoolkit and \bhah, \bah demonstrates that hyperbolic relaxation can be a robust, versatile, and performant approach for AH finding.
\end{abstract}

\maketitle
\sloppy

\mainmatter

\section{Introduction}
\label{sec:intro}

From the first direct detection of gravitational waves (GWs)~\cite{Abbott:2016blz,Abbott:2016izl} to the hundreds of binary black hole (BBH) mergers observed since~\cite{LIGOScientific:2018mvr,LIGOScientific:2020ibl,LIGOScientific:2021usb,KAGRA:2021vkt}, numerical relativity (NR) simulations of black holes (BHs) have formed a cornerstone of GW astrophysics. When a GW event is detected, the observed waveform is compared to \emph{tens of millions} of theoretical predictions to infer the physical parameters of the binary system. These predictions, in turn, rely on NR BBH simulation catalogs as ground truth.

The critical role of NR simulations---and the BHs they model---extends beyond isolated BBH mergers. For instance, binary neutron star (BNS) mergers typically result in a rapidly spinning BH surrounded by a neutron-rich accretion disk. These complex remnants are prime candidates for powering energetic electromagnetic counterparts, including gamma-ray bursts and kilonovae, which are observable from seconds to days after the GW signal. Indeed, the first GW-anchored multimessenger event, GW170817~\cite{GW170817,LIGOScientific:2017ync}, reflects the importance of accurately modeling BHs and BH-forming systems as powerful multimessenger sources---even though the nature of its remnant remains debated.

Together, these examples establish that dynamical BHs---whether merging, accreting, or forming from stellar collapse---play a central role in GW and multimessenger astrophysics. Connecting observational signatures to BH physical parameters necessitates robust methods for identifying and tracking the BHs themselves within NR simulations. This is primarily accomplished using apparent horizon (AH) finding algorithms. An AH corresponds to the outermost marginally outer trapped surface (MOTS) of a BH, defined as a surface on which the expansion of outgoing null geodesics vanishes~\cite{Penrose:1964wq,Senovilla:2011fk,Thornburg:2003sf}. Unlike event horizons, which depend on the entire future spacetime evolution, AHs can be determined from data on individual spatial slices, making them indispensable tools for the real-time identification and characterization of black holes within NR simulations.

In static spacetimes, the AH coincides with the event horizon. In dynamical spacetimes, however, the AH lies strictly inside the event horizon. This property makes it a practical inner boundary for NR simulations, which leverage the AH surface information in several ways.

First, excision boundaries are placed inside the AH, allowing ingoing boundary conditions that exclude BH interiors---including singularities---from the computational domain. Excision is essential for stable BH evolutions in, e.g., the generalized harmonic formulation~\cite{GarfinkleGeneralizedHarmonicpre,PretoriusGeneralizedHarmonic} of GR. Second, informed by AH data, GR hydrodynamic and magnetohydrodynamic evolution schemes can be adjusted to suppress unphysical behavior inside the BH~\cite{Duez:2004uh,Hawke:2005zw,EtienneThirdBHNS,Zilhao:2013hia}, mitigating instabilities and failures in conservative-to-primitive variable recovery.

AHs also enable the extraction of quasi-local BH properties such as mass and spin. These quantities are critical for (i) characterizing merger remnants, (ii) verifying conservation of energy and angular momentum, and (iii) estimating BH parameters when asymptotic quantities are unavailable or unreliable.

For quasi-stationary BHs, the irreducible mass $M_{\rm irr}$ can be estimated from the AH area $A$ as $M_{\rm irr} \approx \sqrt{A/(16\pi)}$, an exact relation in stationary spacetimes. The total (Christodoulou) mass $M$ is computed from $M_{\rm irr}$ and angular momentum $J$, though this requires \emph{a priori} knowledge of $J$~\cite{Christodoulou:1970wf,Alcubierre:2004hr}. Methods to estimate the dimensionless spin parameter $J/M^2$ on a given spatial slice include evaluating the ratio of proper polar and equatorial circumferences of the AH, which is accurate for nearly Kerr geometries~\cite{Smarr:1973zz,Brandt:1994db,Alcubierre:2004hr}. More general methods involve surface integrals of an approximate rotational Killing vector field defined on the AH, as employed by the isolated horizon formalism~\cite{Ashtekar:2004cn}.

The ability to demarcate causally disconnected regions and accurately estimate BH properties reflects the central role AH finders play in NR. AH finding has a long history, spanning decades and encompassing a range of techniques, thoroughly reviewed in Thornburg's 2007 \emph{Living Review}~\cite{Thornburg:2006zb}.

These methods share the common goal of solving the nonlinear elliptic partial differential equation (PDE) for the 2D MOTS in 3D. In spherical symmetry, this equation can be solved via nonlinear root finding, and in axisymmetry with a shooting method. In full 3D, a variety of algorithms are used, with differing levels of robustness, efficiency, and implementation complexity~\cite{Thornburg:2006zb}.

Thornburg's \ahfd~\cite{Thornburg:2003sf}, the most widely used AH finder in the \einsteintoolkit (\et) and possibly the fastest general-purpose implementation, solves the elliptic PDE directly using fourth-order finite differencing on a multi-patch cubed-sphere angular grid. It employs Newton's method with a symbolically differentiated Jacobian to iteratively solve the resulting nonlinear algebraic system. A direct elliptic approach in a similar vein was developed by Fran\c{c}a~\cite{Franca:2023bed} for \grchombo~\cite{Andrade:2021rbd,grchombo_web}, using finite differencing on a spherical grid and \texttt{PETSc} to solve the nonlinear system. Most recently, Hui \& Lin~\cite{Hui:2024ggb} introduced the first multigrid-based AH solver, which required recasting the elliptic PDE into linear and nonlinear components. At very high angular resolutions, their solver outperformed \ahfd, likely due to the superior $\mathcal{O}(N)$ complexity of multigrid methods with increasing grid point counts $N$.

In contrast to direct elliptic solver approaches, \emph{Flow}-based AH finders recast the elliptic equation such that an initial guess surface is evolved forward in a pseudo-time parameter $t$, converging toward a solution of the elliptic PDE as $t \to \infty$. These methods are particularly robust, capable of locating AHs even from poor initial guesses, although they are typically slower than direct elliptic solvers. To the best of our knowledge, all existing flow-based methods recast the elliptic PDE into a \emph{parabolic} equation evolved in pseudo-time~\cite{Thornburg:2006zb}. This includes the \emph{fast}-flow~\cite{Gundlach:1997us} solvers used in \dendrogr~\cite{Fernando:2022php,dendrogr_web}, \spectre~\cite{SpECTRE,spectre_web}, and Alcubierre's \texttt{AHFinder} module in the \et~\cite{Alcubierre:1998rq,EinsteinToolkit}.

This paper introduces \bah, the \bhah AH Algorithm, which departs from traditional flow-based methods by reformulating the elliptic PDE as a \emph{hyperbolic} system and solving it via hyperbolic relaxation. While this approach is not new to NR---having been used in initial-data construction~\cite{Ruter:2017iph,Assumpcao:2021fhq,Tootle:2025ikk} and forming the basis of the widely-used Gamma-driver shift conditions in moving-puncture evolutions~\cite{Alcubierre:2000yz,Campanelli:2005dd}---this work presents its first application to AH finding. Specifically, we recast the elliptic PDE defining a MOTS as a hyperbolic PDE on a sphere, yielding a damped two-dimensional scalar wave equation in which the MOTS equation replaces the Laplacian term.

The hyperbolic system evolves in pseudo-time toward a steady state that satisfies the original elliptic equation. To eliminate spurious behavior on the spherical grid near coordinate singularities at $\theta = 0$ and $\theta = \pi$, we adopt the ``NR in spherical coordinates''~\cite{Brown:2009ki,Montero:2012yr,Baumgarte:2012xy} reference-metric-based approach. This method handles singular parts of tensors analytically, interpolating and finite-differencing only the regular parts. Compared to earlier techniques---such as the cubed-sphere method used in \ahfd~\cite{Thornburg:2003sf}---this approach simplifies the treatment of coordinate singularities while remaining robust to the choice of initial guess, similar to other flow methods.

Another advantage of hyperbolic relaxation techniques is their straightforward integration into existing NR evolution codes, which already include all necessary components for evolving hyperbolic PDEs. Specifically, \bah has been implemented in a single-patch version of the \nrpy-based~\cite{nrpy_web,Ruchlin:2017com} \bhah evolution code, similar to our hyperbolic relaxation solver \nell for NR initial data~\cite{Assumpcao:2021fhq,Tootle:2025ikk}. In contrast, direct elliptic solvers typically require formulating the MOTS PDE as an algebraic system, often involving separate linear algebra packages, initial guesses close to the solution, and nontrivial decompositions of the PDE into linear and nonlinear parts (e.g.,~\cite{Lin:2007cd,Hui:2024ggb}).

\begin{figure}[H]
    \centering
    \includegraphics[width=0.3\textwidth]{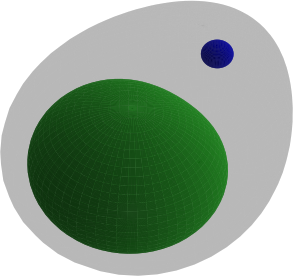}
    \caption{Common apparent horizon (gray) for a $q=4$ binary black hole configuration on a Brill--Lindquist initial data slice, with inner marginally trapped surfaces also plotted (green and blue spheroids). All surfaces were identified by \bah.}
    \label{fig:q4_3d}
\end{figure}

While hyperbolic relaxation methods are generally easier to implement and often more robust than direct elliptic solvers---particularly when given poor initial guesses---they are typically slower. As discussed in Sec.~\ref{subsec:qeq4_bah_innovations}, locating the $q=4$ common horizon shown in Fig.~\ref{fig:q4_3d} from a na\"ive initial guess takes over 10x longer with \bah than with \ahfd. To close this performance gap, we implement \openmp parallelization within \bah and introduce two new techniques within the hyperbolic relaxation framework. First, inspired by multigrid methods, we develop a strategy in which low-cost solutions on coarse grids serve as initial guesses for finer grids, accelerating convergence by more than 12x on 16 cores. Second, we incorporate an \emph{over-relaxation} technique, adapted from numerical linear algebra, to further accelerate convergence. Together, these innovations reduce the runtime for \bah---when locating the challenging $q=4$ common horizon from scratch---to within about 10\% of \ahfd on 16 cores.

In the more common context of dynamically tracking horizons, \bah constructs a high-quality initial guess by extrapolating from up to three previous solutions. This accurate guess offers two key advantages. First, being close to the true solution, it significantly reduces the number of relaxation iterations. Second, it confines the search to a thick spherical shell around the expected horizon location rather than the full spherical volume. Since metric data must be interpolated from the source NR grids onto \bah's spherical grid, reducing the search volume directly lowers the number of expensive interpolations---a major contributor to AH finder runtime. Together, these improvements significantly reduce overhead, enabling \bah to outperform \ahfd by approximately 2.1x in dynamic AH tracking at comparable accuracy (Sec.~\ref{sec:gw150914}).

The remainder of this paper is structured as follows. Section~\ref{sec:basic_eqs} reviews the mathematical foundations underlying \bah, while Sec.~\ref{sec:alg_approach} details its algorithmic implementation. In Sec.~\ref{sec:results}, we present numerical results benchmarking \bah against \ahfd in a range of scenarios, from horizon tracking to highly challenging single-horizon searches. Finally, Sec.~\ref{sec:conclusion} summarizes our findings and potential avenues for future enhancement of \bah.

\section{Basic Equations}
\label{sec:basic_eqs}

We adopt Einstein summation notation, with repeated Latin indices implying summation over the three spatial dimensions. The standard ADM (Arnowitt--Deser--Misner) 3+1 decomposition of the spacetime metric~\cite{Arnowitt:1962hi} is given by
\beq
ds^2 = -\alpha^2 dt^2 + \gamma_{ij} (dx^i + \beta^i dt)(dx^j + \beta^j dt),
\eeq
where $\alpha$ is the lapse function, $\beta^i$ the shift vector, and $\gamma_{ij}$ the spatial 3-metric. This decomposition naturally introduces spatial hypersurfaces characterized by an extrinsic curvature $K_{ij}$, defined by
\beq
K_{ij} = -\frac{1}{2\alpha}(\partial_t \gamma_{ij} - D_i \beta_j - D_j \beta_i),
\eeq
where $D_i$ is the covariant derivative compatible with $\gamma_{ij}$.

\subsection{Expansion Function \texorpdfstring{$\Theta$}{Theta}}

An AH is defined as the \emph{outermost} marginally outer trapped surface (MOTS) of a BH. Formally, a MOTS appears when the expansion function $\Theta$, which measures the rate of change of the area of an infinitesimal 2-surface along its outward normal, vanishes:
\beq
\Theta = D_i s^i - K + s^i s^j K_{ij} = 0,
\label{eq:Theta}
\eeq
where $s^i$ is the outward unit normal to the MOTS, $K_{ij}$ is the extrinsic curvature of the spatial hypersurface, and $K = \gamma^{ij} K_{ij}$ is its trace.

In practice, the MOTS is defined implicitly by a scalar level-set function $F(r,\theta,\phi)$ given in spherical coordinates as
\beq
F(r,\theta,\phi)=r-h(\theta,\phi)=0,
\label{eq:F}
\eeq
so that $F<0$ denotes the interior and $F>0$ the exterior of the MOTS. Since
\beq
D_i F = \partial_i F = \lambda\, s_i \quad \mbox{with} \quad \lambda=\sqrt{\gamma^{ij}\partial_iF\,\partial_jF},
\label{eq:lambda}
\eeq
the unit normal is obtained as
\beq
s^i = \frac{\gamma^{ij}\partial_jF}{\lambda}.
\label{eq:sUi}
\eeq
The full covariant divergence of $s^i$ is given by
\beq
D_i s^i = \frac{1}{\sqrt{\gamma}}\partial_i\Bigl(\sqrt{\gamma}\,s^i\Bigr)
= \partial_i s^i + s^i\,\partial_i\Bigl(\ln\sqrt{\gamma}\Bigr),
\eeq
where the partial divergence $\partial_i s^i$ can be computed from Eq.~\ref{eq:sUi} as
\beq
\partial_i s^i = \frac{1}{\lambda}\Bigl(\partial_j F\,\partial_i \gamma^{ij}+\gamma^{ij}\partial_i\partial_jF\Bigr)
-\frac{1}{\lambda^2}\partial_i\lambda\,\gamma^{ij}\partial_jF,
\eeq
so that the full covariant divergence becomes
\beq
D_i s^i = \frac{1}{\lambda}\Bigl(\partial_j F\,\partial_i \gamma^{ij}+\gamma^{ij}\partial_i\partial_jF\Bigr)
-\frac{1}{\lambda^2}\partial_i\lambda\,\gamma^{ij}\partial_jF
+ s^i\,\partial_i\Bigl(\ln\sqrt{\gamma}\Bigr).
\eeq
Combining these results, the expansion function is expressed as
\beq
\fbox{$\displaystyle
        \begin{array}{l}
            \Theta = \frac{1}{\lambda} \left( \partial_j F\, \partial_i \gamma^{ij} + \gamma^{ij} \partial_i \partial_j F \right)
            - \frac{1}{\lambda^2} \partial_i \lambda\, \gamma^{ij} \partial_j F + s^i\, \partial_i \left( \ln \sqrt{\gamma} \right) \\
            \quad\quad -\, K + s^i s^j K_{ij} = 0
        \end{array}$}
\label{eq:Theta_expanded}
\eeq
which, given the definition of $F$ (Eq.~\ref{eq:F}), $\lambda$ (Eq.~\ref{eq:lambda}), and $s^i$ (Eq.~\ref{eq:sUi}), along with the 3-metric $\gamma_{ij}$, its derivatives, and the extrinsic curvature $K_{ij}$, constitutes a nonlinear elliptic PDE for $h(\theta,\phi)$.

\subsection{Hyperbolic Relaxation Equations}
\label{subsec:hyperbolic_relax_eqs}

Having cast the expansion function $\Theta$ as a nonlinear elliptic PDE for $h(\theta,\phi)$ (Eq.~\ref{eq:Theta_expanded}), we solve it using a hyperbolic relaxation scheme~\cite{Ruter:2017iph,Assumpcao:2021fhq}. The method evolves the system in a pseudo-time $t$ via the following damped wave equations:

\beq
\begin{array}{rl}
    \partial_t h & = v - \eta h,  \\[6pt]
    \partial_t v & = -c^2 \Theta,
\end{array}
\label{eq:evolution_eqs}
\eeq
where $\eta$ is a damping coefficient (units $1/M$) and $c$ is the relaxation wave speed (dimensionless, set to 1 in \bah). In the limit $t\to \infty$, the damping drives the system toward a steady state with $\partial_t v = 0$, recovering $\Theta = 0$.

This method preserves the full nonlinearity of $\Theta$, avoiding the need for linearization or high-quality initial guesses, and is thus robust in practice. It also integrates seamlessly into existing hyperbolic evolution frameworks used in NR, such as \bhah (generated by \nrpy~\cite{nrpy_web,Ruchlin:2017com}).

\subsection{Proper Area-Based MOTS Diagnostics}

Integration over the MOTS is central to AH diagnostics. The infinitesimal area element $dS$ is determined by the induced metric $q_{\alpha\beta}$ (where Greek indices run over angular coordinates $\theta, \phi$), computed from an embedding function that describes the surface in 3D Cartesian coordinates $p^k(\theta,\phi)$. For a MOTS given by $r = h(\theta,\phi)$ in spherical coordinates $(r, \theta, \phi)$ centered at a Cartesian origin $x^i_o$, the Cartesian coordinates $p^k(\theta,\phi)$ of points on this surface are obtained via the standard spherical-to-Cartesian transformation (see Eqs.~\ref{eq:xh_x}--\ref{eq:xh_z}, with $x^i_o$ as the origin for the $h(\theta,\phi)$ radial definition). The area element is then:
\[
    dS = \sqrt{q}\,d\theta\,d\phi,
\]
where $q$ is the determinant of the induced metric.

The induced metric $q_{\alpha\beta}$ is derived by pulling back the ambient 3D metric $\gamma_{kl}$ onto the 2D MOTS using the tangent vectors to the surface, $\partial p^k / \partial y^\alpha$, where $y^\alpha = (\theta, \phi)$:
\beq
q_{\alpha\beta} = \frac{\partial p^k}{\partial y^\alpha} \frac{\partial p^l}{\partial y^\beta} \gamma_{kl}.
\eeq

Explicitly substituting $p^k(\theta,\phi)$ and the derivatives of $h(\theta,\phi)$ yields an expression for $\sqrt{q}$ (cf. Eq.~(24) in \cite{Hui:2024ggb}):
\beqn
dS & = & \biggl[ \bigl(\gamma_{rr}h_{,\theta}^2 + 2\gamma_{r\theta}h_{,\theta} + \gamma_{\theta\theta}\bigr)
    \bigl(\gamma_{rr}h_{,\phi}^2 + 2\gamma_{r\phi}h_{,\phi} + \gamma_{\phi\phi}\bigr) \nonumber \\
    &   & \quad - \bigl(\gamma_{rr}h_{,\theta}h_{,\phi} + \gamma_{r\theta}h_{,\phi} + \gamma_{r\phi}h_{,\theta}
    + \gamma_{\theta\phi}\bigr)^2 \biggr]^{1/2} d\theta\,d\phi.
\label{eq:dS_explicit}
\eeqn
Here, the components $\gamma_{ij}$ are evaluated in spherical coordinates at $r = h(\theta, \phi)$ (Eq.~\ref{eq:F}).

From this expression, the total MOTS area $A$ and irreducible mass $M_{\rm irr}$ are computed as:\footnote{Strictly speaking, the second equality holds only in stationary spacetimes, but we follow convention in \bah and define the irreducible mass accordingly.}
\beq
A = \int_0^\pi d\theta \int_{-\pi}^{\pi} d\phi\, \sqrt{q}, \quad M_{\rm irr}=\sqrt{\frac{A}{16\pi}}.
\eeq

To update the MOTS location in \bah, we compute the centroid $x^i_{\rm c}$ as the area-weighted average of the Cartesian position vector $x^i_h(\theta, \phi)$:
\beq
x^i_{\rm c} = \frac{1}{A} \int_0^\pi d\theta \int_{-\pi}^\pi d\phi\, x^i_h(\theta, \phi) \sqrt{q}.
\label{eq:centroid}
\eeq
Using spherical coordinates centered at $x^i_o = (x^x_o, x^y_o, x^z_o)$, the components of $x^i_h$ are given by the standard transformation:
\begin{eqnarray}
    x^x_h(\theta, \phi) &= x^x_o + h(\theta, \phi) \sin\theta \cos\phi, \label{eq:xh_x} \\
    x^y_h(\theta, \phi) &= x^y_o + h(\theta, \phi) \sin\theta \sin\phi, \label{eq:xh_y} \\
    x^z_h(\theta, \phi) &= x^z_o + h(\theta, \phi) \cos\theta. \label{eq:xh_z}
\end{eqnarray}

We also compute the minimum, maximum, and mean radii relative to the centroid. The mean radius $\langle r \rangle$ is defined as the area-weighted average of the Euclidean distance from the centroid:
\beq
\langle r\rangle = \frac{1}{A} \int_0^\pi d\theta \int_{-\pi}^\pi d\phi\, \sqrt{\delta_{kl}(x^k_h(\theta, \phi) - x^k_c)(x^l_h(\theta, \phi) - x^l_c)}\,\sqrt{q},
\eeq
where $\delta_{kl}$ is the Kronecker delta.

\subsection{Proper Circumference-Based MOTS Diagnostics}

The spin of an isolated, nearly stationary BH can be estimated by computing proper circumferences along coordinate planes on the MOTS~\cite{Smarr:1973zz}. We define these via path integrals:
\begin{eqnarray*}
    C_{\rm xy} &=& \oint_{\phi=-\pi}^{\pi} \sqrt{q_{\phi\phi}(\theta=\pi/2)}\, d\phi, \\
    C_{\rm xz} &=& \int_{\mathrm{meridional\ loop\ at\ }\phi=0} \sqrt{q_{\theta\theta}}\, d\ell_\theta, \\
    C_{\rm yz} &=& \int_{\mathrm{meridional\ loop\ at\ }\phi=\pi/2} \sqrt{q_{\theta\theta}}\, d\ell_\theta,
\end{eqnarray*}
where $d\ell_\theta$ denotes integration along a closed meridional path of constant $\phi$, running from pole to pole and back along the MOTS. For numerical evaluation, the line element is interpolated along the path and integrated using high-order quadrature.

For spin aligned with the $z$-axis, the dimensionless spin magnitude $\chi^z = a^z/M$ is determined from the polar-to-equatorial circumference ratio $C_{\rm r} = C_{\rm p}/C_{\rm e}$, with $C_{\rm p} = C_{\rm xz} = C_{\rm yz}$ and $C_{\rm e} = C_{\rm xy}$. Following Alcubierre et al.~\cite{Alcubierre:2004hr}, we invert
\beq
C_{\rm r} = \frac{1+\sqrt{1-(\chi^z)^2}}{\pi}\,E\left(-\frac{(\chi^z)^2}{[1+\sqrt{1-(\chi^z)^2}]^2}\right),
\label{eq:c_r_spin}
\eeq
where $E(k)$ is the complete elliptic integral of the second kind:
\[
    E(k) = \int_0^{\pi/2} \sqrt{1 - k \sin^2\theta}\, d\theta.
\]
The inversion is performed via the Newton-Raphson algorithm, requiring both $E(k)$ and its derivative, which depends on the complete elliptic integral of the first kind, $K(k)$.

Because Eq.~\ref{eq:c_r_spin} depends only on $(\chi^z)^2$, it yields only the magnitude $|\chi^z|$, not its sign. To estimate $|\chi^x|$ and $|\chi^y|$, we substitute $\chi^z \rightarrow \chi^x$ or $\chi^y$ into Eq.~\ref{eq:c_r_spin} and apply it to different coordinate planes. Specifically, we use the ratios $C_{\rm xz}/C_{\rm yz}$ and $C_{\rm xy}/C_{\rm yz}$ for $|\chi^x|$, and $C_{\rm xy}/C_{\rm xz}$ and $C_{\rm yz}/C_{\rm xz}$ for $|\chi^y|$.

\section{Algorithmic Approach}
\label{sec:alg_approach}

Like its sister code \nell~\cite{Assumpcao:2021fhq}, \bah is a simplified variant of the \bhah~\cite{Etienne:2024ncu} NR evolution code, generated by the open-source \nrpy~\cite{nrpy_web,Ruchlin:2017com} framework.\footnote{The version of \bah used in this paper can be generated from a Python environment, for example via the Linux command line, using \texttt{pip install git+https://github.com/nrpy/nrpy.git@92a51d8 \&\& python3 -m nrpy.examples.bhahaha}, and following the emitted instructions.} \nrpy leverages \SymPy~\cite{sympy} for symbolic computation and significantly extends its code generation capabilities, converting complex tensorial expressions (e.g., the expansion $\Theta$, Eq.~\ref{eq:Theta_expanded}) into highly optimized C code. \bah is implemented as a standalone AH finder library---the first of its kind---intended for general use in \emph{any} NR code.\footnote{Instructions for implementing \bah into NR codes are provided in the \nrpy repository, where \bah is housed.}

The following sections outline the core steps of the \bah algorithm, detailing how it acquires and processes spacetime metric information from the host NR code, locates an AH using hyperbolic relaxation, and computes diagnostics on it.

\subsection{Reading and Processing Metric Data}
\label{subsec:data_preprocessing}

\bah requires the host NR code to provide the spatial metric, $\gamma_{ij}$, and the extrinsic curvature, $K_{ij}$, in a Cartesian basis, interpolated onto \bah's 3D cell-centered spherical grid. The grid's origin must lie within the AH and fully enclose it. The grid may be configured either as a full sphere---for initial horizon finding---or as a spherical shell of nonzero thickness for subsequent tracking.

Although \bah seeks a 2D surface defined by $r = h(\theta, \phi)$, evaluating the expansion $\Theta$ requires metric variables and their spatial derivatives at every point on the evolving trial surface in 3D. Performing full 3D interpolations from the host NR code onto $r = h(\theta,\phi)$ at each relaxation step would be computationally prohibitive. To avoid this, \bah precomputes all necessary geometric quantities on a 3D spherical grid that shares the same angular resolution as $h$. This arrangement enables rapid \emph{1D interpolation along radial spokes} (i.e., in $r$ at fixed $\theta,\phi$) to obtain metric data on the trial surface throughout the relaxation process.

After the host NR code interpolates the 12 independent ADM quantities ($\gamma_{ij}$ and $K_{ij}$) onto \bah's grid, \bah transforms them into conformal BSSN variables expressed in the spherical basis. BSSN variables are used instead of ADM largely for convenience, as they are already implemented within the reference-metric formalism in \nrpy. To ensure numerical stability, the singular parts of tensor components in this formalism are handled analytically, so that only the regular (nonsingular) parts are ever interpolated or finite-differenced~\cite{Montero:2012yr,Baumgarte:2012xy,Ruchlin:2017com}.

To compute $\Theta$, all required BSSN variables and first spatial derivatives must be available for interpolation onto the surface $r = h(\theta,\phi)$ at each pseudo-time step. These derivatives (both angular and radial) are computed on the 3D grid using sixth-order finite differencing, employing appropriate boundary conditions (for angular derivatives: parity conditions at poles and $\phi=\pm\pi$; for radial derivatives upwind/downwind stencils at radial boundaries). These precomputed fields are stored on the 3D spherical grid, ready for 1D radial interpolations onto the trial AH surface.

\subsection{Hyperbolic Relaxation Formulation}
\label{subsec:hyperbolic_relaxation}

As introduced in Sec.~\ref{subsec:hyperbolic_relax_eqs}, \bah finds a marginally outer trapped surface (MOTS) by evolving the hyperbolic relaxation system defined in Eq.~\ref{eq:evolution_eqs}. In these equations, the function $h(\theta,\phi)$ represents the trial surface radius on \bah's spherical grid, while the auxiliary field $v$ serves as a radial velocity-like component during the pseudo-time evolution. For simplicity, the relaxation wave speed $c$ is set to unity ($c = 1$) in this implementation.\footnote{In \nell~\cite{Assumpcao:2021fhq}, setting $c$ to the maximum value allowed by the CFL condition significantly accelerated convergence on prolate spheroidal grids. This was effective largely because of the extreme \emph{radial} grid spacing variations on those highly distorted coordinates. While the CFL condition for \bah's spherical grid is still set by the small azimuthal spacing near the poles, the use of a cell-centered \emph{strictly angular} grid keeps this minimum spacing manageable and the overall variation less severe than in the prolate spheroidal case. As a result, we found this optimization to be ineffective here.}

The damping term $-\eta h$ in the equation for $\partial_t h$ drives the system toward the steady-state condition $v = \eta h$. As the system relaxes and $\partial_t v$ also approaches zero, the second equation enforces the desired MOTS condition $\Theta = 0$.

To ensure dimensional consistency and achieve scale-invariant behavior independent of the BH's coordinate mass, the user provides a characteristic BH mass scale $m_{\rm scale}$ and specifies the dimensionless damping coefficient $m_{\rm scale}\eta$. Internally, \bah computes $\eta = (m_{\rm scale}\eta)/m_{\rm scale}$ so that the damping term $\eta h$ carries the correct physical units.

\subsection{Initial Data}
\label{subsec:initial_data}

When no prior guess is provided, \bah employs a default spherical guess $h = 0.8 R_{\rm search}$, where $R_{\rm search}$ is the user-specified maximum radius of the spherical search volume. For subsequent finds (tracking), \bah constructs an initial guess at $t_N$ by extrapolating data from up to three previous horizon configurations. Lagrange extrapolation of up to second order (depending on the number of available prior configurations) is consistently applied to \emph{all quantities}. This algorithm predicts: (i) the new centroid $\vec{x}_c(t_N)$ from prior centroid locations $\vec{x}_c(t_{N-k})$; (ii) the expected radial extent $[r_{\min}^{\rm src}(t_N), r_{\max}^{\rm src}(t_N)]$ from prior radial bounds; and (iii) the horizon shape function $h(t_N, \theta,\phi)$ pointwise from the prior shapes $h(t_{N-k}, \theta,\phi)$. The predicted radial extent defines a thick spherical shell onto which the host code interpolates metric data, significantly reducing expensive interpolations compared to using a full spherical volume. The extrapolated shape $h(t_N, \theta,\phi)$ serves as the initial guess, denoted $h(t=0, \theta, \phi)$, for the pseudo-time evolution described by Eq.~\ref{eq:evolution_eqs}.

\subsection{Pseudo-Time Evolution}
\label{subsec:pseudotime_evol}

The system Eq.~\ref{eq:evolution_eqs} is integrated forward in pseudo-time using the standard third-order Strong Stability Preserving Runge-Kutta (SSPRK3) scheme~\cite{GottliebShuTadmorSSPRK}. The pseudo-time step $\Delta t$ is determined by a Courant-Friedrichs-Lewy (CFL) condition based on the wave speed $c = 1$ and the proper grid spacings on the evolving surface, $\Delta s_\theta$ and $\Delta s_\phi$. SSPRK3 was selected after empirical testing showed it permitted a slightly higher CFL factor (1.0) than other methods considered, without increasing computational cost.

Evaluating the right-hand side of Eq.~\ref{eq:evolution_eqs}, specifically the $\Theta$ term, requires the precomputed metric quantities and their derivatives (Sec.~\ref{subsec:data_preprocessing}) on the current trial surface $r = h(t,\theta,\phi)$. Since this surface generally does not align with the radial grid points of \bah's internal 3D grid, these quantities are obtained via interpolation. A key optimization is performed: at the start of each full pseudo-time step (before the first SSPRK3 substep), \bah performs high-order 1D Lagrange interpolation along radial spokes for each angular grid point $(\theta_j,\phi_k)$, mapping the required precomputed fields from the 3D grid onto the trial surface $r = h(\theta_j,\phi_k)$.

These interpolated values are cached and reused during all subsequent SSPRK3 substeps within that pseudo-time step. This strategy dramatically reduces computational cost compared to repeated 3D interpolations, while maintaining high-order accuracy. Only the angular derivatives of the evolving shape function $h(\theta,\phi)$ (needed for $\Theta$) are recomputed at each SSPRK3 substep using high-order finite differencing, as $h$ itself changes within the Runge-Kutta integration.

Ghost zones for the evolving fields $h(\theta,\phi)$ and $v(\theta,\phi)$ are updated after each SSPRK3 substep using the same parity-based method~\cite{Ruchlin:2017com,Assumpcao:2021fhq} as used for metric quantities (Sec.~\ref{subsec:data_preprocessing}).

\subsection{Stop Conditions}
\label{subsec:stop_conditions}

This pseudo-time evolution proceeds until the trial surface $h(\theta,\phi)$ sufficiently satisfies the condition $\Theta = 0$. To determine convergence, \bah employs physically meaningful and scale-invariant criteria.

Because $\Theta$ has dimensions of inverse mass (or inverse length in $G=c=1$ units), requiring the $L_\infty$ norm of $\Theta$ to be less than $10^{-5}$ would be far more stringent for a BH with coordinate mass $1$ than for one with mass $10^{-3}$. To ensure consistent convergence behavior across a broad range of systems---including high mass ratio binaries---\bah requires the user to specify a characteristic mass scale $m_{\rm scale}$ for each BH. All convergence criteria are then defined in terms of the dimensionless product $m_{\rm scale} \Theta$.

To our knowledge, \bah is the first general-purpose AH finder to explicitly incorporate $\Theta$'s dimensionality into its stopping conditions, a critical step for ensuring robustness regardless of arbitrary overall mass scale or mass ratio.

Convergence is declared and relaxation terminated when both of the following criteria are satisfied:
\beq
||m_{\rm scale} \Theta||_\infty < \epsilon_\infty \quad \mathrm{and} \quad ||m_{\rm scale} \Theta||_2 < \epsilon_2,
\label{eq:stop_criteria}
\eeq
where $||\cdot||_\infty$ and $||\cdot||_2$ denote the $L_\infty$ and $L_2$ norm over the angular grid, respectively. The default tolerances are $\epsilon_\infty = 10^{-5}$ and $\epsilon_2 = 10^{-2}$. This is consistent with the general preference to adopt $L_\infty$ norms as thresholds for convergence. This said, the tolerances can be adjusted to prioritize averaged accuracy if desired. In typical scenarios, we have never observed $L_\infty$ and $L_2$ norms to differ by more than a factor of a few. By comparison \ahfd adopts $\epsilon_\infty = 10^{-8}$, but as shown in Sec.~\ref{sec:gw150914}, the extra orders of magnitude offer no benefit for typical horizon finds.

\subsection{Convergence Acceleration Techniques}
\label{subsec:acceleration}

Although hyperbolic relaxation is robust, its convergence is limited by the CFL condition, with computational cost scaling as roughly $\mathcal{O}(N^{d+1})$ for a $d$-dimensional surface ($d=2$ here) discretized with $N$ points per dimension. To mitigate this, \bah typically operates at modest angular resolution (defaulting to $N_\theta \times N_\phi = 32 \times 64$), made effective by its use of high-order numerical methods (sixth-order finite differencing, Sec.~\ref{subsec:data_preprocessing}, and fifth-order Lagrange interpolation). As shown in Sec.~\ref{sec:results}, \bah achieves comparable accuracy to the fourth-order \ahfd~\cite{Thornburg:2003sf} at their respective default resolutions and tolerances. To further accelerate convergence towards the state defined by the termination criteria (Sec.~\ref{subsec:stop_conditions}), \bah introduces two techniques new to hyperbolic relaxation solvers:

\subsubsection{Multigrid-Inspired Relaxation}
\label{subsubsec:multigrid}

\bah implements a relaxation strategy inspired by multigrid methods, but tailored to hyperbolic evolution. Rather than transferring residuals between resolution levels as in standard multigrid elliptic solvers, \bah focuses on generating progressively better initial guesses. It first relaxes the system on a coarse angular grid (e.g., $8\times16$ points), where convergence is rapid due to the small number of grid points and the relatively large pseudo-time step permitted by the CFL condition. The resulting solution $h(\theta, \phi)$ is then prolonged to the next finer grid (e.g., $16\times32$) via fifth-order Lagrange interpolation and used as an initial guess\footnote{Generally, high-order Lagrange prolongation will introduce high-frequency noise to the solution, which may slow convergence at higher resolutions. Future work will explore Hermite or bicubic spline interpolation in this step.}. This hierarchy may continue up to the target resolution (e.g., $32\times64$), with each finer level requiring significantly fewer iterations due to the improved starting point. The 3D cell-centered spherical source grid containing BSSN data (Sec.~\ref{subsec:data_preprocessing}) is prepared once per resolution level. These data are initially interpolated from the host code at the highest angular resolution; for coarser levels, the required 3D BSSN fields (including derivatives) are downsampled (interpolated) versions of high-resolution data.

\subsubsection{Over-Relaxation}
\label{subsubsec:overrelaxation}
During pseudo-time evolution, the system's early wavelike behavior often transitions to a slower, asymptotic convergence. To accelerate this phase, \bah periodically attempts over-relaxation at intervals proportional to the horizon light-crossing time, using a stored prior shape $h_p$ (at pseudo-time $t_p$) and the current shape $h_c$ (at $t_c$). The algorithm searches for an optimal extrapolation factor, $\alpha_{\rm opt}$, by generating a sequence of trial shapes
\[
    h_{\rm trial}(\alpha) = h_p + \alpha\, (h_c - h_p).
\]
The search begins with $\alpha=2.0$ and tests a geometric progression of values (increasing by a factor of 1.2), computing the residual norm $||m_{\rm scale}\Theta||_\infty$ for each. The process identifies the factor that minimizes this norm and terminates as soon as the norm ceases to decrease. If an optimal factor is found that yields a relative improvement in the norm greater than a threshold of $5\times 10^{-5}$, the extrapolation is accepted: the shape is updated to $h_{\rm trial}(\alpha_{\rm opt})$, the auxiliary velocity field is reset to $v=\eta h$ to damp the dynamics, metric data are re-interpolated onto the new surface, and the CFL-limited timestep is recalculated. The stored history ($h_p, t_p$) is then cleared. If an over-relaxation step is not accepted, the current state is saved as the new history for a future attempt.

\subsection{Diagnostics Computation}
\label{subsec:diagnostics}

Upon successful convergence (i.e., when criteria Eq.~\ref{eq:stop_criteria} are satisfied), \bah computes several diagnostic quantities to provide key physical parameters characterizing the BH at that moment in time, as well as verify the accuracy of the horizon found by \bah (i.e., norms of $\Theta$):
\begin{itemize}
    \item \textbf{Area \& Mass:} The proper surface area $A$ is computed by integrating the area element $dS = \sqrt{q}\, d\theta\, d\phi$ over the surface $r=h(\theta,\phi)$, where $q$ is the determinant of the induced 2-metric on the horizon (Eq.~\ref{eq:dS_explicit}). The irreducible mass is then estimated as $M_{\rm irr} = \sqrt{A/(16\pi)}$ (exact for stationary horizons).
    \item \textbf{Centroid \& Radii:} The coordinate centroid $x^i_c$ of the horizon surface is computed using Eq.~\ref{eq:centroid}. The minimum, maximum, and mean coordinate radii relative to this centroid are also reported.
    \item \textbf{Spin (via Circumferences):} The proper equatorial circumferences in the coordinate planes passing through the centroid ($C_{\rm xy}, C_{\rm xz}, C_{\rm yz}$) are computed. These are used to estimate the magnitude of the dimensionless spin components $|\chi^i|$ by numerically inverting the Kerr spin-circumference relations given in Eq.~\ref{eq:c_r_spin}.
    \item \textbf{Expansion Norms:} The final $L_\infty$ and $L_2$ norms of the dimensionless expansion $m_{\rm scale}\Theta$ are used to determine whether termination criteria were met.
\end{itemize}

\section{Results}
\label{sec:results}

We compare \bah's performance, robustness, and accuracy with the widely used \ahfd across several challenging scenarios that showcase its key innovations and utility as a general-purpose AH finder. Section~\ref{subsubsec:resultsbah_accelerations} evaluates the performance gains from \bah's acceleration strategies and \openmp parallelization using a demanding $q=4$ common horizon search with a poor initial guess. Section~\ref{subsubsec:results_scale_invariance} tests the scale invariance of \bah and \ahfd by repeating this search across BH total masses spanning 16 orders of magnitude. While such single-slice tests serve as useful benchmarks, AH finders are more commonly used for tracking horizons in dynamical spacetimes. To this end, Sec.~\ref{sec:gw150914} assesses \bah's performance during a BBH inspiral and merger, identifying the tolerances required to match typical \ahfd errors. Finally, Sec.~\ref{subsec:three_bh_crit} shows that at high resolutions \bah matches \ahfd and other finders to many significant digits in a precision 3-BH common horizon search, further demonstrating its infrastructure independence through implementation in the \bhah NR code.

\subsection{\bah vs. \ahfd Performance Comparison: \texorpdfstring{$q=4$}{q=4} BBH Merger}
\label{subsec:qeq4_bah_innovations}

To assess the acceleration techniques of Sec.~\ref{subsec:acceleration}, the \et implementation of \bah is used to locate both individual and common MOTSs on a $q=4$ two-BH zero-momentum  Brill--Lindquist~\cite{Brill:1963yv} spacetime slice with bare masses $m_1=0.2$ and $m_2=0.8$, placed at $(0.05,0.05,0.05)$ and $(-0.235,-0.235,-0.235)$, respectively. This setup yields the individual and common MOTSs plotted in Fig.~\ref{fig:q4_3d}.

The configuration is made intentionally challenging for AH finders. The large mass ratio creates a highly non-spherical common horizon, and the axis of symmetry is tilted ${\sim}\,45^\circ$ from \bah's $z$-axis. Both \bah and \ahfd are given the same poor initial guesses: three spherical trial surfaces centered on the massive puncture (radius 0.4), the less massive one (radius 0.1), and the common horizon (radius 0.9), with the latter centered at the origin $(0,0,0)$---far from its true location.

\subsubsection{Impact of Multigrid and Over-Relaxation Accelerations on \bah Performance Compared to \ahfd}
\label{subsubsec:resultsbah_accelerations}

\begin{figure}[ht]
    \centering
    \includegraphics[width=0.9\textwidth] {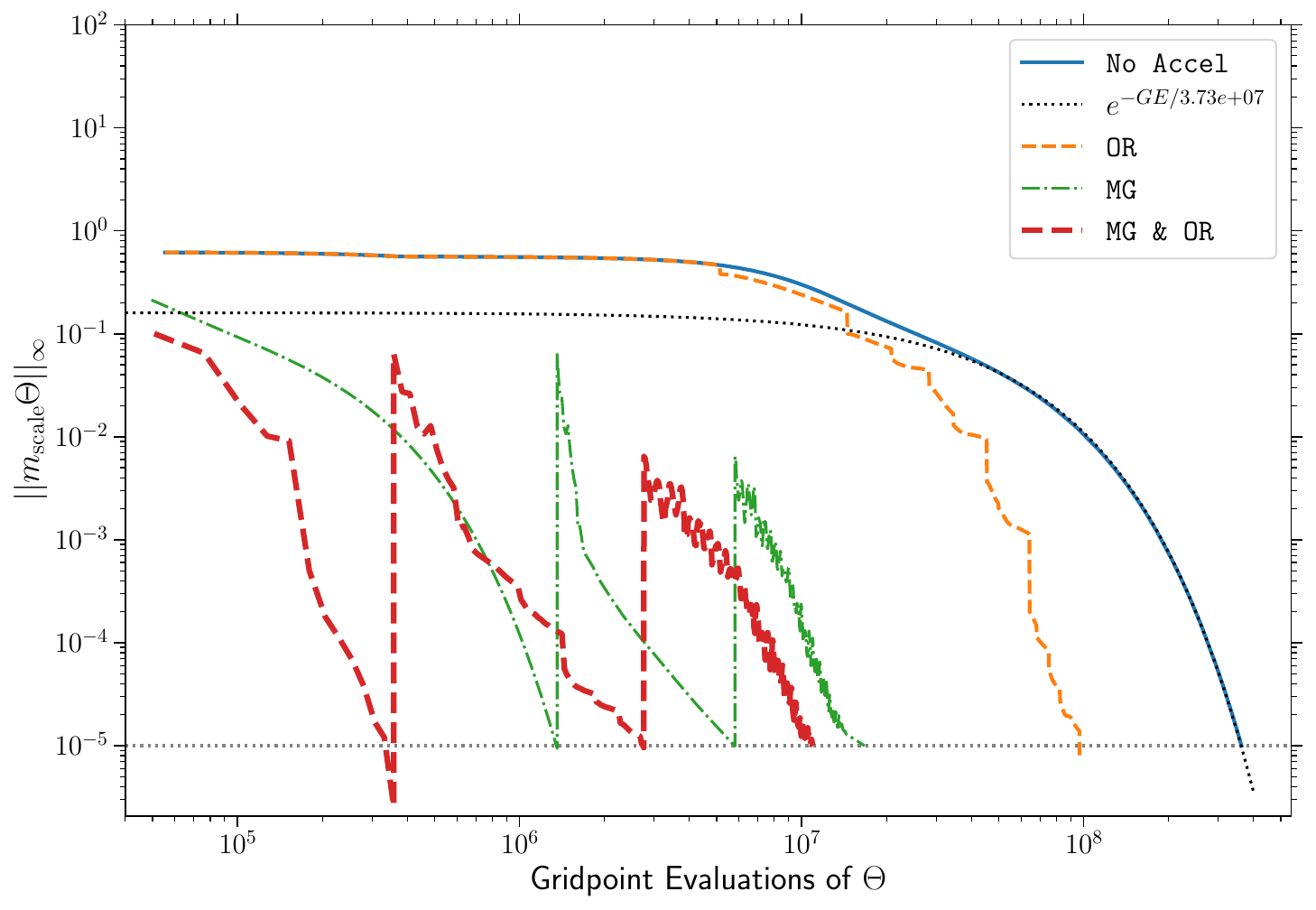}
    \caption{\textbf{Acceleration Study.} Convergence of the $L_\infty$ norm of the dimensionless expansion ($||m_{\rm scale}\Theta||_{\infty}$) versus the total number of gridpoint evaluations of $\Theta$ (GEs) for the $q=4$ common horizon search starting from a poor initial guess located at the origin. Comparison of: baseline hyperbolic relaxation (\texttt{No Accel}, solid blue); an exponential fit to its late-time behavior (\texttt{$e^{-{\rm GE}/(3.73 \times 10^7)}$}, dotted black); relaxation using only the multigrid-inspired strategy (\texttt{MG}, dash-dot green); using only over-relaxation (\texttt{OR}, dashed orange); and using the combination of both techniques (\texttt{MG \& OR}, thick dashed red). The horizontal dotted line indicates the target convergence threshold ($||m_{\rm scale}\Theta||_{\infty} {\approx}\, 10^{-5}$).}
    \label{fig:accel_study}
\end{figure}

We establish a baseline by comparing the performance of unmodified \bah to that of \ahfd in locating the $q=4$ common horizon, using the same poor, origin-centered spherical initial guess. At the default resolution ($32\times64$), unaccelerated \bah is significantly slower than the serial \ahfd,\footnote{\ahfd, as implemented in the \et, supports parallelism across multiple initial guesses but is not \openmp-parallelized for a single search. Since our benchmarks involve one initial guess at a time, this parallelism is not used here.} requiring 10--52x more wall-clock time depending on CPU core count (Table~\ref{tab:timing_results}, first row). Its convergence from the poor initial guess (Fig.~\ref{fig:accel_study}, solid blue line) begins with wave-like transients and variable damping, followed by steady exponential decay. Fitting this late-time decay (dotted black line) gives an e-folding period of $3.73\times10^7$ total gridpoint evaluations of $\Theta$ (GEs). Reaching the target tolerance $||m_{\rm scale}\Theta||_{\infty}=10^{-5}$ (horizontal dotted line) requires approximately $5\times10^8$ GEs.

\begin{table}[h]
    \caption{\label{tab:timing_results}Timing benchmarks (in seconds) comparing \bah and \ahfd for the $q=4$ test case. We highlight the impact of \bah acceleration techniques---over-relaxation (OR), multigrid-inspired (MG), and their combination---as well as its scalability with CPU core count. \ahfd is not \openmp-parallelized; its single-core time (1.41\,s) serves as the primary reference, with multi-core timings showing similar performance. \textit{Note:} Although \ahfd lacks \openmp parallelization, it \emph{is} parallelized across multiple simultaneous horizon solves (multiple initial guesses) as described in Appendix~B of Thornburg~\cite{Thornburg:2003sf}. Our single-horizon benchmarks use one initial guess per solve, so this inter-solve parallelism is not exercised. All results are averaged over three runs on a 16-core AMD Ryzen 9 5950X CPU. For clarity, the primary comparison between fully accelerated \bah and \ahfd is shown in boldface.}
    \begin{indented} 
        \item[]\begin{tabular}{@{}llrrrrr}
            \hline
                          &              & \multicolumn{5}{c}{Number of Cores}                                                     \\
            \cline{3-7}
                          &              & 1                                   & 2          & 4          & 8          & 16         \\
            \hline
            \bah no accel & total        & 73.0                                & 37.6       & 21.1       & 14.6       & 20.6       \\
            \hline
            \bah OR       & total        & 19.9                                & 10.7       & 6.06       & 4.43       & 5.81       \\
            \hline
            \bah MG       & total        & 4.01                                & 2.47       & 1.71       & 1.38       & 1.70       \\
            \hline
            \bah MG \& OR & 3D interp    & 0.54                                & 0.60       & 0.66       & 0.67       & 0.69       \\
                          & horizon find & 2.28                                & 1.22       & 0.68       & 0.47       & 0.65       \\
                          & total        & {\bf 2.82}                          & {\bf 1.82} & {\bf 1.34} & {\bf 1.14} & {\bf 1.34} \\
            \hline
            \ahfd         & total        & {\bf 1.41}                          & {\bf 1.41} & {\bf 1.44} & {\bf 1.44} & {\bf 1.48} \\
            \hline
        \end{tabular}
    \end{indented}
\end{table}

The multigrid-inspired strategy (Sec.~\ref{subsubsec:multigrid}) substantially boosts performance. It solves the problem on successively finer grids---$8 \times 16$, $16 \times 32$, and $32 \times 64$---with each solution initializing the next. This hierarchy accelerates convergence, as achieving convergence on coarse grids is inexpensive: just $1/64$ and $1/8$ of the GEs per iteration compared to $32 \times 64$. As shown in Fig.~\ref{fig:accel_study} (green dash-dot line), the method achieves a $10^{-5}$ tolerance using only ${\sim}\, 1/33$ of the GEs required by unaccelerated \bah. Wall-clock time improves by 11--18x across 1--16 cores (Table~\ref{tab:timing_results}), cutting single-core runtime from 73.0\,s to 4.01\,s. Jagged features in the convergence plot mark grid resolution transitions.

The over-relaxation technique alone (Sec.~\ref{subsubsec:overrelaxation}) provides a modest 3.3--3.7x speedup across 1--16 cores (Fig.~\ref{fig:accel_study}, orange dashed line; Table~\ref{tab:timing_results}), reducing single-core runtime from 73.0\,s to 19.9\,s. This approach is generally less effective at improving performance than the multigrid-inspired strategy.

Combining periodic over-relaxation with multigrid yields the highest performance boost (Fig.~\ref{fig:accel_study}, thick dashed red line), achieving a $10^{-5}$ tolerance in ${\sim}\, 8 \times 10^6$ GEs. This strategy offers a 13--26x speedup across 1--16 cores, lowering the single-core runtime to 2.82\,s (Table~\ref{tab:timing_results}).

Comparing parallel scaling of fully accelerated \bah (`\bah MG \& OR') to serial \ahfd yields additional insight. On one core, \bah takes 2.82\,s---twice the time of \ahfd (1.41\,s)---but with the aid of \openmp scaling is able to reduce runtime to 1.14\,s on 8 cores. This outperforms \ahfd (1.44\,s) by 21$\%$. At 16 cores, \bah slows to 1.34\,s, slightly above its 8-core time, likely due to increased \openmp overhead due to overdecomposition on the $32 \times 64$ grid.

A breakdown of \bah's timing (`\bah MG \& OR', Table~\ref{tab:timing_results}) better illustrates its scaling with increased core count. While the `horizon find' computation scales well to 8 cores, the `3D interp' phase---interpolating from the \et \Carpet AMR grid~\cite{Carpet,CarpetCode:web} to \bah's spherical grid---dominates runtime and scales poorly, limiting overall speedup.

\subsubsection{Scale Invariance: \bah vs. \ahfd}
\label{subsubsec:results_scale_invariance}

Since Einstein's equations in vacuum are scale-invariant, robustness to total mass scale is a key test of AH finders: rescaling mass ($M \to \lambda M$) and coordinates ($x^\alpha \to \lambda x^\alpha$) maps solutions to other valid ones. Thus dimensionless quantities like $A/M^2$ should remain constant. We test this using $q=4$ nonspinning Brill--Lindquist BBH data (Sec.~\ref{subsec:qeq4_bah_innovations}), rescaling the puncture masses $m_1, m_2$ so that $M=m_1+m_2$ spans $10^{-8}$ to $10^8$. Grid spacing is scaled proportionally ($h \propto M$) to keep $h/M$ fixed.

At each scale, both \bah and \ahfd find the common horizon and inner MOTSs. We compute the normalized areas $A_c/M^2$, $A_1/M^2$ (larger mass), and $A_2/M^2$ (smaller mass). \bah uses a scale-invariant convergence criterion $||m_{\rm scale}\Theta||_\infty=10^{-5}$, while \ahfd uses an absolute tolerance $||\Theta||_\infty=10^{-5}$ chosen for consistency.

As shown in Table~\ref{tab:mass_area}, the finders agree well across $10^{-3} \le M \le 10^3$: $A_c/M^2$ and $A_1/M^2$ match to five digits; $A_2/M^2$ to four. Outside this range, \bah maintains six-digit accuracy across all 16 orders of magnitude, confirming scale invariance. \ahfd, however, fails for $M \le 10^{-5}$ (missing horizons) and diverges for $M \ge 10^5$, with $A_c/M^2$ deviating by up to 18\% at $M=10^8$. This suggests \ahfd's absolute tolerance may cause false convergence at large $M$ due to premature settling near the initial guess.

\begin{table}[ht]
    \centering
    \caption{Scale invariance study comparing \bah (\texttt{BAH}) and \ahfd (\texttt{AHFD}) for locating MOTSs in a $q=4$ BBH initial dataset. The table lists the normalized areas of the common horizon ($A_c/M^2$), larger inner surface ($A_1/M^2$), and smaller inner surface ($A_2/M^2$) versus total mass scale $M$, with grid resolution $h/M$ fixed. NF (Not Found) indicates failure to locate the surface.}
    \label{tab:mass_area}
    \begin{tabular}{c|cc|cc|cc}
        \hline
        Mass Scale ($M$) & \multicolumn{2}{c|}{Common ($A_c/M^2$)} & \multicolumn{2}{c|}{Inner 1 ($A_1/M^2$)} & \multicolumn{2}{c}{Inner 2 ($A_2/M^2$)}                                                                        \\
        \cline{2-7}
                         & \texttt{BAH}                            & \texttt{AHFD}                            & \texttt{BAH}                            & \texttt{AHFD}             & \texttt{BAH} & \texttt{AHFD}             \\
        \hline
        $10^{-8}$        & 50.1715                                 & \textbf{NF}                              & 46.5112                                 & \textbf{NF}               & 6.58904      & \textbf{NF}               \\
        $10^{-5}$        & 50.1715                                 & 50.171\textbf{3}                         & 46.5112                                 & 46.511\textbf{1}          & 6.58904      & \textbf{NF}               \\
        $10^{-3}$        & 50.1715                                 & 50.171\textbf{3}                         & 46.5112                                 & 46.511\textbf{1}          & 6.58904      & 6.589\textbf{38}          \\
        1                & 50.1715                                 & 50.171\textbf{3}                         & 46.5112                                 & 46.511\textbf{1}          & 6.58904      & 6.589\textbf{38}          \\
        $10^{3}$         & 50.1715                                 & 50.171\textbf{3}                         & 46.5112                                 & 46.511\textbf{1}          & 6.58904      & 6.589\textbf{38}          \\
        $10^{5}$         & 50.1715                                 & 5\textbf{8}.\textbf{9557}                & 46.5112                                 & 46.51\textbf{92}          & 6.58904      & \textbf{7}.\textbf{47378} \\
        $10^{8}$         & 50.1715                                 & 5\textbf{8}.\textbf{9557}                & 46.5112                                 & 4\textbf{7}.\textbf{0723} & 6.58904      & \textbf{7}.\textbf{30382} \\
        \hline
    \end{tabular}
\end{table}

\subsection{Dynamical Horizon Study: GW150914}
\label{sec:gw150914}

Having established \bah's robustness to challenging horizon finds on a single spacetime slice, we next evaluate its performance in a fully dynamical setting: an NR simulation of the GW150914-like BBH merger from the \et gallery~\cite{wardell_barry_2016_155394,EinsteinToolkit:web}.

The simulation makes use of the best-fit priors from GW150914~\cite{Abbott:2016blz,TheLIGOScientific:2016wfe} with a mass ratio $q = 36/29 {\approx}\, 1.24$, aligned spins $\vec\chi_1=\{0,0,0.31\}$ and $\vec\chi_2=\{0,0,-0.46\}$, and initial separation $D=10M$, with total mass $M = m_1 + m_2$. The initial data is computed using \et's \texttt{TwoPunctures} thorn~\cite{AnsorgTwoPunctures,BaikalVacuum2024}, with a resolution of ($48\times48\times20$) to minimize initial constraint violations. The gauge is initialized with $\alpha = e^{-2\phi}$ and $\beta^i = 0$.

The evolution uses the BSSN formalism implemented in \BaikalVacuum~\cite{Ruchlin:2017com,BaikalVacuum2024} with the moving-puncture gauge (1+log lapse, Gamma-driver shift). Numerical methods include 8th-order spatial differencing, RK4 time integration, and 9th-order Kreiss--Oliger dissipation (strength 0.15). Punctures are tracked using the \texttt{PunctureTracker} thorn and AMR is managed by \Carpet~\cite{Carpet,CarpetCode:web} using 11 refinement levels (factor 2) and an outer boundary of $1365.\bar{3}M$. The finest resolution is $\Delta x_{\rm finest} = M/36 \approx 0.0278M$. Subcycled time integration uses a CFL factor of 0.45, with timestep ratios: $\{1, 1, 1, 1, 1, 2, 4, 8, 16, 32, 32\}$. Full parameters may be found in the \texttt{ET\_BHaHAHA} thorn.\footnote{The \texttt{ET\_BHaHAHA} thorn used in this paper is available in the \texttt{EinsteinAnalysis} repository at \url{https://bitbucket.org/einsteintoolkit/einsteinanalysis}, commit ID \texttt{b306cb}. The parameter file used for this example can be found at \texttt{ET\_BHaHAHA/par/GW150914\_baikalvacuum.par}.}

To make the testbed more challenging, we adopt standard moving puncture methods, deliberately omitting recent noise-suppression techniques~\cite{Etienne:2024ncu} that can reduce constraint violations on AMR grids in BBH simulations by orders of magnitude. Both \bah and the standard \ahfd~\cite{Thornburg:2003sf} concurrently track AHs, searching every 44 base iterations ($\Delta t_{\rm search}\approx0.55M$). Initially, each identifies two separate horizons.

\begin{figure}[ht]
    \centering
    \includegraphics[width=0.75\textwidth]{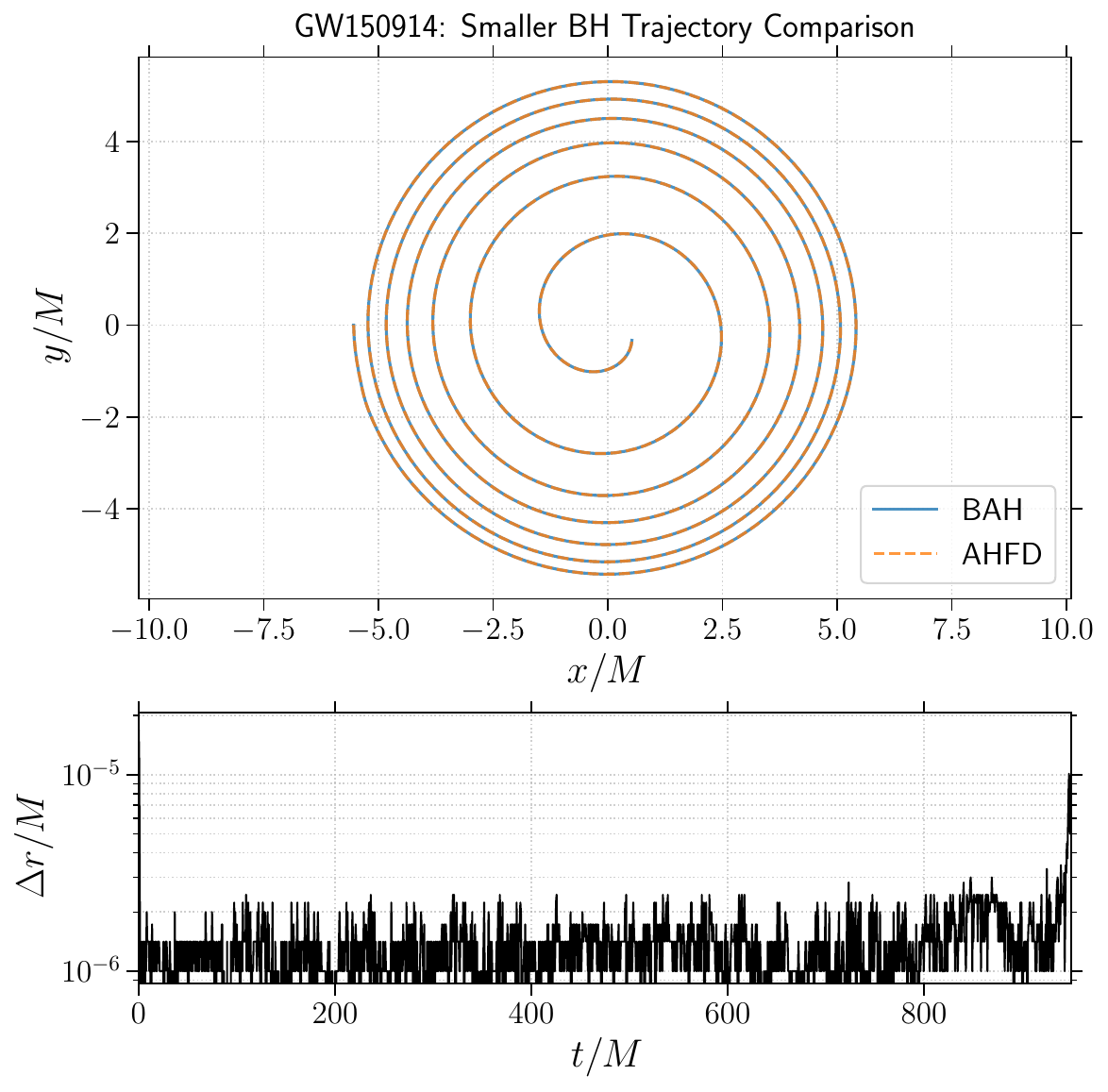}
    \caption{Top panel: Trajectory of the centroid of the smaller BH apparent horizon during the inspiral for GW150914, as tracked by \bah (BAH, solid blue line) and \ahfd (AHFD, dashed orange line). Coordinates are scaled by the total mass $M$. Bottom panel: Absolute Euclidean distance $\Delta r/M$ between the centroids found by the two methods at corresponding simulation iterations, plotted against time $t/M$.}
    \label{fig:bbh_trajectory}
\end{figure}

Figure~\ref{fig:bbh_trajectory} illustrates the inspiral dynamics. The top panel shows the smaller black hole's AH centroid trajectory as computed independently by \bah and \ahfd during the ${\sim}\,6.5$ orbits preceding common horizon formation. The trajectories are visually indistinguishable, indicating strong agreement. The bottom panel quantifies this, showing their separation (dimensionless Euclidean distance) $\Delta r/M$ remains near $10^{-6}$, confirming the high precision of both finders.

For dynamic tracking, \bah employs three multigrid levels ($8\times16$, $16\times32$, $32\times64$), over-relaxation, quadratic extrapolation from up to three prior solutions to estimate the initial shape and centroid, and optimized spherical shell search domains to reduce interpolation costs from the host NR code.

While these features substantially improve \bah's performance, they must be evaluated against accuracy constraints: tolerances required to ensure agreement with \ahfd. As we will show, interpolating metric quantities ($\gamma_{ij}$ and $K_{ij}$) from the \et AMR grid introduces a systematic error. Both finders rely on this interpolation to compute derivatives used in evaluating the expansion $\Theta$, leading to an error floor of approximately $10^{-5}$.

\begin{figure}[ht]
    \centering
    \includegraphics[width=0.9\textwidth]{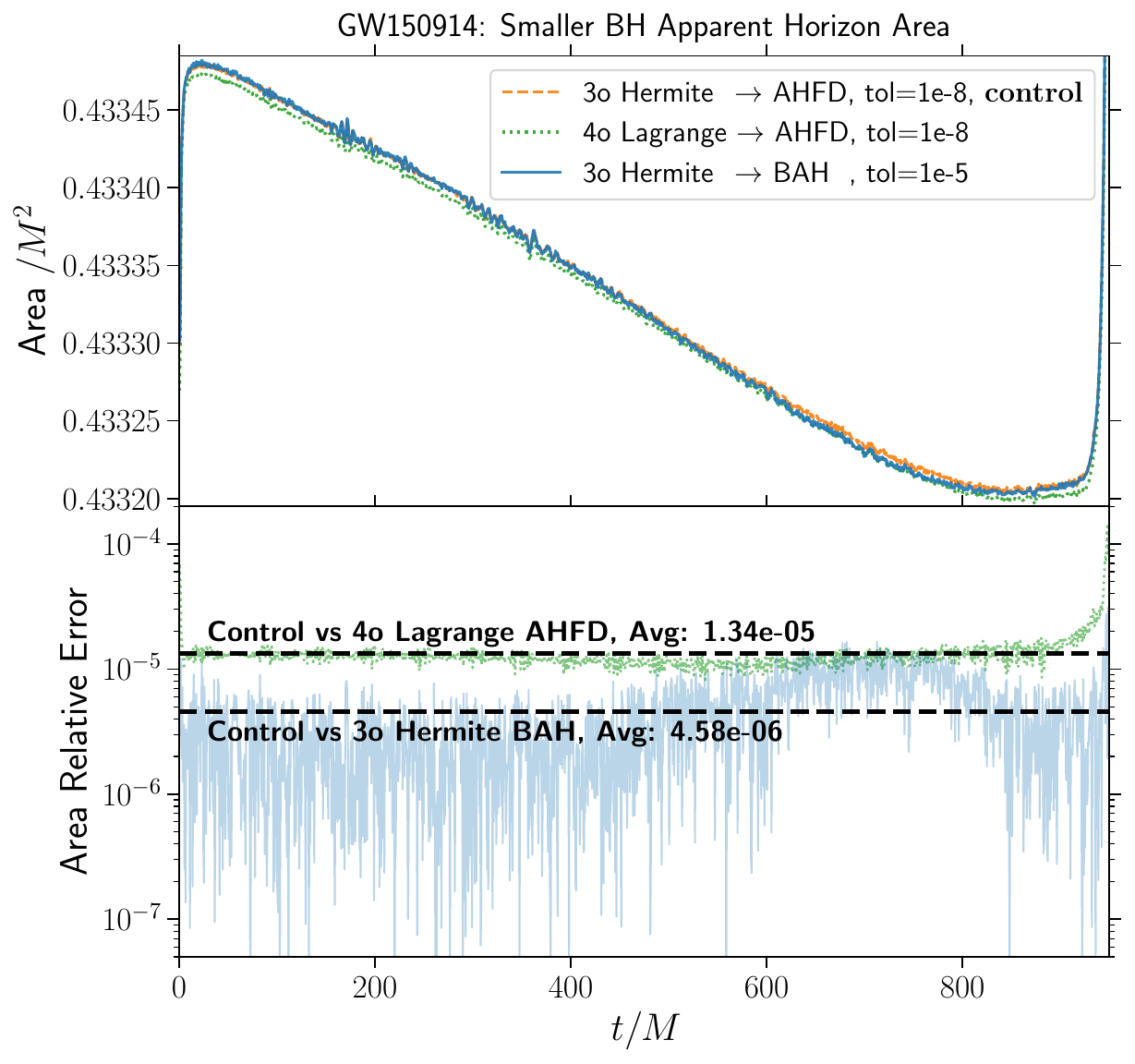}
    \caption{AH area comparison for the smaller BH in the GW150914 simulation. Top panel: Area evolution over time for three finder configurations: the control run using \ahfd with 3rd-order Hermite interpolation ($10^{-8}$ tolerance; dashed orange), \ahfd using 4th-order Lagrange interpolation ($10^{-8}$ tolerance; dotted green), and \bah (3rd-order Hermite interpolation, $10^{-5}$ tolerance; solid blue). Bottom panel: Absolute relative area difference compared to the control run's AHFD area. The average relative error for \bah ($10^{-5}$ tol.) vs control is $\approx 4.58 \times 10^{-6}$ (blue line, lower dashed black average line). The average relative error for \ahfd (4th-order Lagrange) vs control is $\approx 1.34 \times 10^{-5}$ (green dotted line, upper dashed black average line).}
    \label{fig:bbh_area_comparison}
\end{figure}

To measure the impact of interpolation error from the host NR code, in Fig.~\ref{fig:bbh_area_comparison}, we plot the area of the smaller horizon throughout the entire inspiral using three different approaches. First, \ahfd at default $||\Theta||_\infty<10^{-8}$ tolerance using the \et's 3rd-order Hermite interpolation algorithm (\textbf{control}). Second, an identically configured \ahfd, except instead of Hermite, the \et's 4th-order Lagrange interpolation algorithm is used. Third, \bah is run with its default tolerance of $m_{\rm scale}||\Theta||_\infty=10^{-5}$.

As shown in the bottom panel of Fig.~\ref{fig:bbh_area_comparison}, comparing \ahfd with Hermite versus Lagrange interpolation reveals an average relative area difference of $1.34 \times 10^{-5}$. This difference, consistent with the general accuracy of ${\sim}\, 10^{-5}$ reported in the abstract of the \ahfd announcement paper~\cite{Thornburg:2003sf}, establishes an acceptable relative error floor for our comparisons.

At its default tolerance and using the \et's 3rd-order Hermite interpolation, \bah demonstrates excellent agreement with the control run (top panel, Fig.~\ref{fig:bbh_area_comparison}). Its average relative area error of $4.58 \times 10^{-6}$ is comfortably below the $1.34 \times 10^{-5}$ interpolation error floor.

Applying the same error analysis approach illustrated in Fig.~\ref{fig:bbh_area_comparison}, we conduct a detailed error analysis on both $L_2$ and $L_\infty$ norms of $\Theta$ in Tables~\ref{tab:bah_l2_errors} and~\ref{tab:linf_errors}.

\begin{table}[ht]
    \centering
    \caption{\bah (BAH) $L_2$ Tolerance Study: average relative apparent-horizon area errors for varying BAH $L_2$ tolerances, compared to the control setup---\ahfd (AHFD) with 3rd-order Hermite interpolation and default tolerance $||\Theta||_\infty<10^{-8}$. The top row isolates interpolation error on \Carpet AMR grids by comparing 3rd-order Hermite and 4th-order Lagrange interpolants at fixed default tolerance. To examine the impact of varying BAH $L_2$ tolerance alone, the $L_\infty$ tolerance was fixed at $8\times10^{-2}$, which is typically satisfied early; convergence requires both tolerances to be met. The final column lists total runtimes in BAH and AHFD during inspiral. For AHFD, this includes time spent in \texttt{find\_horizons()} and \texttt{store()}, excluding excision-mask setup (not applicable to BAH). The top-row runtime reflects only the AHFD control case with 3rd-order Hermite interpolation.}
    \label{tab:bah_l2_errors}
    \begin{tabular}{@{}l c c@{}}
        \hline
        Description                                        & $E_{\rm rel,\ control}$ & Runtime (s) \\
        \hline
        Input Metric Interp. Error (AHFD)                  & $1.34 \times 10^{-5}$   & 500.        \\
        BAH $m_{\rm scale}||\Theta||_2= 2 \times 10^{-5}$  & $8.68 \times 10^{-6}$   & 201         \\
        BAH $m_{\rm scale}||\Theta||_2 = 10^{-5}$          & $5.18 \times 10^{-6}$   & 204         \\
        BAH $m_{\rm scale}||\Theta||_2 = 2 \times 10^{-6}$ & $4.05 \times 10^{-6}$   & 256         \\
        BAH $m_{\rm scale}||\Theta||_2 = 8 \times 10^{-7}$ & $4.16 \times 10^{-6}$   & 293         \\
        \hline
    \end{tabular}
\end{table}

\begin{table}[ht]
    \centering
    \caption{\bah (BAH) and \ahfd (AHFD) $L_\infty$ Tolerance Study: average relative AH area errors across various nominal $L_\infty$ tolerances, compared to the same control configuration as in Table~\ref{tab:bah_l2_errors}. The top row isolates interpolation error by comparing 3rd-order Hermite and 4th-order Lagrange methods on \Carpet AMR grids, with fixed default tolerance. BAH uses per-horizon scale-invariant tolerances $m_{\rm scale}||\Theta||$ with $m_{\rm scale}\approx0.45$ and $0.55$ (bare puncture masses), while AHFD uses a dimensionful $||\Theta||$ scaled by 0.5 (average mass). To isolate the effect of BAH's $L_\infty$ tolerance, the $L_2$ tolerance was fixed at $8\times10^{-2}$ and is generally met early; both must be satisfied for convergence. The final column lists total runtimes for BAH and AHFD during inspiral, as detailed in Table~\ref{tab:bah_l2_errors}. For the top row, we report only AHFD control runtime with 3rd-order Hermite interpolation.}
    \label{tab:linf_errors}
    \begin{tabular}{@{}l c c c@{}}
        \hline
                                                             & AHFD Error                & BAH Error                 & AHFD; BAH   \\
        Description                                          & ($E_{\rm rel,\ control}$) & ($E_{\rm rel,\ control}$) & Runtime (s) \\
        \hline
        Input Metric Interp. Error (AHFD)                    & $1.34 \times 10^{-5}$     & N/A                       & 500.; ---   \\
        AHFD \& BAH $m_{\rm scale}||\Theta||_\infty=10^{-4}$ & $3.53 \times 10^{-8}$     & $2.09 \times 10^{-5}$     & 359; 193    \\
        AHFD \& BAH $m_{\rm scale}||\Theta||_\infty=10^{-5}$ & $7.49 \times 10^{-11}$    & $4.58 \times 10^{-6}$     & 494; 240    \\
        AHFD \& BAH $m_{\rm scale}||\Theta||_\infty=10^{-6}$ & $5.87 \times 10^{-12}$    & $4.15 \times 10^{-6}$     & 494; 333    \\
        AHFD \& BAH $m_{\rm scale}||\Theta||_\infty=10^{-7}$ & $1.34 \times 10^{-12}$    & $4.10 \times 10^{-6}$     & 497; 559    \\
        AHFD \& BAH $m_{\rm scale}||\Theta||_\infty=10^{-8}$ & $2.67 \times 10^{-12}$    & $4.08 \times 10^{-6}$     & 501; 870    \\
        \hline
    \end{tabular}
\end{table}

Tightening \bah's $L_2$ tolerance ($m_{\rm scale}||\Theta||_2$), while keeping $L_\infty$ loose, reduces the relative area error until it saturates at ${\sim}\,4.16 \times 10^{-6}$ for $m_{\rm scale}||\Theta||_2 = 8\times 10^{-7}$ (Table~\ref{tab:bah_l2_errors}). For \ahfd, reducing the $L_\infty$ tolerance improves precision to $2.67 \times 10^{-12}$ at $||\Theta||_\infty = 10^{-8}$ (Table~\ref{tab:linf_errors}), but this is artificial, as it relies on identical metric data (provided via 3rd-order Hermite interpolation) from the host NR grid. Similarly, \bah's error plateaus at ${\sim}\,4\times 10^{-6}$ when only the $L_\infty$ constraint is tightened.

Comparing the convergence of $L_2$ and $L_\infty$ norms (Tables~\ref{tab:bah_l2_errors},~\ref{tab:linf_errors}), both saturate around $4 \times 10^{-6}$, limited by interpolation from the NR host grid. However, $L_2$ reaches this limit more efficiently. For instance, \bah achieves $4.05 \times 10^{-6}$ in 256s with $m_{\rm scale}||\Theta||_2 = 2 \times 10^{-6}$, while $L_\infty$ needs $m_{\rm scale}||\Theta||_\infty = 10^{-6}$ and 333s to reach $4.15 \times 10^{-6}$.

At practical tolerances ($10^{-5}$), \bah($L_2$) reaches interpolation-limited accuracy faster than \bah($L_\infty$): $5.18\times10^{-6}$ in 204\,s vs $4.58\times10^{-6}$ in 240\,s, both well below the interpolation-error floor. Either norm suffices at ${\sim}10^{-5}$, whereas \ahfd is often configured with much stricter $L_\infty$ tolerances (${\sim}10^{-8}$).

To remain consistent with the stop conditions used by most existing AH finders, we adopt $L_\infty$ as the primary convergence criterion in this work, with default tolerances $m_{\rm scale} ||\Theta||_\infty = 10^{-5}$ and $m_{\rm scale} ||\Theta||_2 = 10^{-2}$. This effectively disables the $L_2$ condition in favor of $L_\infty$. While $L_\infty$ enforces strict local accuracy, $L_2$ better optimizes average accuracy and is less sensitive to outliers. As demonstrated in Tables~\ref{tab:bah_l2_errors} and~\ref{tab:linf_errors}, comparable accuracy can be achieved with lower computational cost by reversing these roles: setting $m_{\rm scale} ||\Theta||_2 = 10^{-5}$ as the primary condition and relaxing the $L_\infty$ constraint to $m_{\rm scale} ||\Theta||_\infty = 10^{-2}$. We therefore recommend this configuration for dynamic horizon tracking scenarios.

At $10^{-5}$ tolerances, \bah achieves similar accuracy in less time than \ahfd. In GW150914-like inspirals run on a single AMD EPYC 9654 (96-core) node using 8 MPI ranks and 12 \openmp threads per rank, \bah achieves interpolation-limited accuracy ($||\Theta|| {\sim}\, 10^{-5}$) in 204--240 seconds, compared to ${\sim}\,$494 seconds for \ahfd at similar error levels. At stricter tolerances ($10^{-7}$--$10^{-8}$), beyond the interpolation limit, \ahfd becomes more efficient as \bah saturates. The scaling studies presented in Sec.~\ref{subsubsec:resultsbah_accelerations} and Table~\ref{tab:timing_results} show that \bah performs optimally at around 8--16 \openmp threads. Thus, the 12-thread configuration used here is nearly ideal. Even with fewer threads, however, \bah would likely retain a substantial speed advantage over \ahfd at standard $10^{-5}$ simulation tolerances.

\bah's efficiency at practical tolerances has important implications for large-scale simulations. Since each MPI rank performs one horizon search, the per-node speedup translates directly into reduced overall cost. For standard BBH evolutions at $10^{-5}$ tolerance, \bah reduces horizon-finding wall time by roughly a factor of two compared to \ahfd, without compromising accuracy. This speedup is easily achieved by enabling the \bah thorn and choosing appropriate tolerances, with no other changes to the existing \et workflow.

\subsubsection{Automatic Triggering of Common Horizon Searches: \bah's BBH Mode}

\ahfd requires specifying in advance when and where to search for a common horizon---a nontrivial task, especially for BBH mergers occurring away from the origin. To address this, \bah's \et implementation includes an experimental ``BBH mode'' that automatically schedules common horizon searches based on the centroids and radii of individual AHs.

\begin{figure}[ht]
    \centering
    \includegraphics[width=0.3\textwidth]{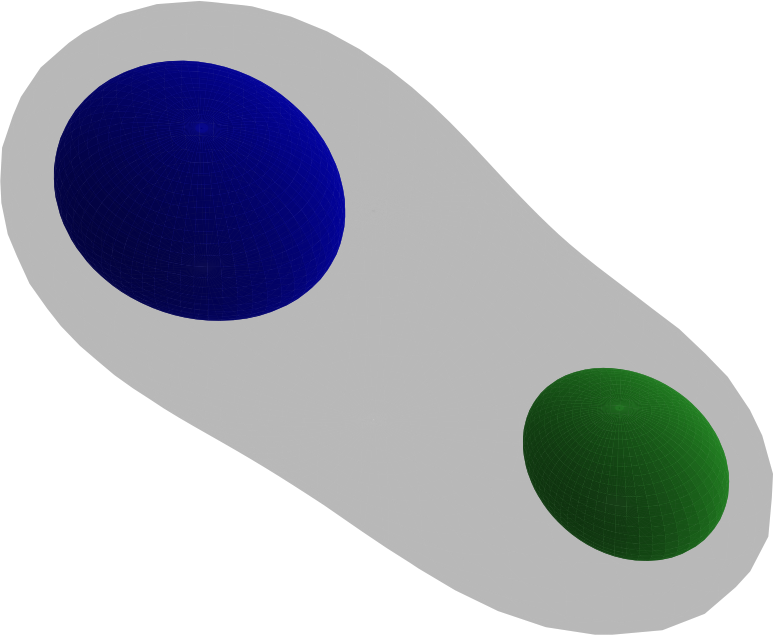}
    \caption{GW150914-like BBH: First common horizon (gray) found, with inner marginally trapped surfaces plotted (green and blue spheroids). All surfaces identified by \bah.}
    \label{fig:gw150914_common_horizon_detail}
\end{figure}

In this GW150914-inspired simulation, ``BBH mode'' scheduled a common-horizon search at iteration 75680 ($t \approx 946M$), and \bah subsequently detected the common horizon shown in Fig.~\ref{fig:gw150914_common_horizon_detail} at iteration 75944 ($t \approx 950M$). For direct comparison, \ahfd, manually triggered to begin common-horizon searches at iteration 75680 as well, also first detected the common horizon at exactly iteration 75944.

To improve robustness during the highly dynamical early phase of a just-formed common horizon, ``BBH mode'' doubles the multigrid resolution at the lowest refinement level, raises the iteration ceiling, reduces the CFL factor by 10\%, and uses a large fixed-radius initial guess for the first three horizon searches.

Although currently experimental, the implementation of ``BBH mode'' within the \et serves as a template for integrating it into other NR codes. Future work includes porting it to the \bhah NR code and enhancing its robustness across a wider range of BBH scenarios.

\paragraph*{Relation to horizon pretracking.}
Schnetter, Herrmann, and Pollney introduced \emph{horizon pretracking} as a complementary strategy for anticipating the formation of a common apparent horizon in BBH simulations~\cite{Schnetter:2004mc,Thornburg:2006zb}. Rather than waiting for a MOTS with $\Theta=0$ to exist, pretracking follows families of (modified) constant-expansion surfaces---e.g., targets defined by $\Theta=\mathrm{const}$, $r\,\Theta=\mathrm{const}$, or $r^2\Theta=\mathrm{const}$---throughout the inspiral. As the binary tightens, the tracked surface's expansion approaches zero, yielding (i) an \emph{early} and \emph{localized} prediction for the time and place of common-horizon formation, and (ii) a high-quality initial guess for a standard elliptic AH solve the instant $\Theta\!\to\!0$. Operationally, a pretracking update typically evaluates several surfaces per pretracking iteration (each trial solve comparable in cost to a single AH solve) and is run intermittently---often every ${\sim}10$ evolution steps rather than every step---so the overhead is adjustable~\cite{Schnetter:2004mc}. Beyond timing, pretracking can provide predictive geometric information (e.g., approximate area/shape) that may inform gauge or excision choices.

Our BBH mode addresses a different operational need: low-overhead, \emph{just-in-time} automation of common-horizon searches within an existing AH-tracking workflow. It uses diagnostics from the individual horizons (centroids and radii) to schedule common-horizon solves and to seed them with extrapolated shapes/centroids, optionally tightening numerical controls immediately after formation. In short, pretracking provides predictive, pre-merger diagnostics~\cite{Schnetter:2004mc}, whereas BBH mode provides robust triggering and seeding at minor additional cost; the two are thus complementary. We recommend pretracking when advance warning and predictive geometric information are desired (e.g., to inform gauge/excision choices), and BBH mode when simplicity and throughput are paramount within a production workflow.

\subsection{Precision Horizon Finding: Three-BH Critical Radius}
\label{subsec:three_bh_crit}

Thus far, we have demonstrated that results from \bah and \ahfd generally agree to within 5--7 significant digits (Secs.~\ref{subsubsec:results_scale_invariance} and~\ref{sec:gw150914}), a level primarily limited by metric-interpolation errors from the \et's Cartesian grids. To investigate whether \bah can achieve even higher accuracy and match other AH finders's results at greater precision, we focus here on a more demanding benchmark.

This test also serves to demonstrate \bah's infrastructure independence. Here, we use its implementation within the \bhah NR code~\cite{Ruchlin:2017com,Etienne:2024ncu,nrpy_web}, rather than the \et~\cite{EinsteinToolkit,EinsteinToolkit:web} framework used previously. Notably, \bhah employs 8th-order finite differencing and 7th-order interpolation, compared to 6th- and 5th-order schemes in \et. This higher-order approach and optimized grid support improved precision.

\begin{figure}[ht]
    \centering
    \includegraphics[width=0.4\textwidth]{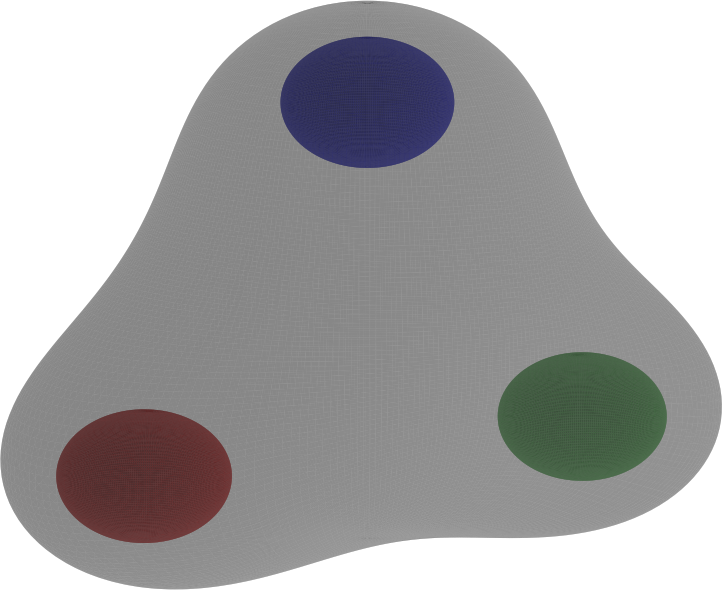}
    \caption{Common apparent horizon (gray) for three equal-mass ($m=1$) Brill-Lindquist black holes, with inner horizons also shown (blue, red, green). The image depicts the system at the largest separation $R$ for which a common horizon was identified (i.e., $R=1.1954995582$) at a convergence tolerance of $m_{\rm scale}||\Theta||_\infty = 3 \times 10^{-8}$ (for total mass $m_{\rm scale}=3$). All horizons were found using \bah implemented within the \bhah framework.}
    \label{fig:three_bh_common_horizon}
\end{figure}

The benchmark problem is the three-black-hole ``critical-radius'' configuration described by Thornburg~\cite{Thornburg:2003sf}. As shown in Fig.~\ref{fig:three_bh_common_horizon}, three equal-mass Brill--Lindquist punctures (each with bare mass $m=1$, total mass $M=3$) are placed in an equilateral triangle centered at the origin. The task is to find the largest separation $R$ admitting a common AH. This critical value, $R_{\rm crit}$, is a longstanding test for AH finders~\cite{Thornburg:2003sf,Hui:2024ggb}.

To isolate \bah's intrinsic accuracy and reduce metric-interpolation error, spacetime data are sampled from a high-resolution \texttt{SinhSpherical} grid native to \bhah. This grid uses a uniform angular resolution of $N_{\theta}\times N_{\phi}=128\times256$ and 320 exponentially spaced radial points,\footnote{Radial points follow $r(x_0) = r_{{\rm max}} \frac{\sinh(x_0/w)}{\sinh(1/w)}$~\cite{Ruchlin:2017com}, with $r_{{\rm max}} = 10$, $w = 0.4$.} efficiently resolving the smooth tri-nodal horizon. Internally, \bah was tested at increasing multigrid resolutions ($64\times128$, $96\times192$, $128\times256$) with a strict convergence criterion: $m_{\rm scale}||\Theta||_\infty < 3\times10^{-8}$, where $m_{\rm scale}=3$ for this system.

Using a bisection-style search,\footnote{The search sampled $12$ radii uniformly between $0.9923076923$ and $2.0076923076$ (code units). At each step, $12$ trial radii were evaluated in parallel on six nodes (two per node).} $R_{\rm crit}$ was bracketed to about 1 part in $10^9$:
\[
    1.1954995582 < R_{\rm crit} < 1.1954995597.
\]
This lies well within the Richardson-extrapolated value $R_{\rm crit} = 1.19549953 \pm 5$ (error bar corresponds roughly to 1 part in $10^7$) from Thornburg using \ahfd~\cite{Thornburg:2003sf}, and matches the recent $1.1954995$ reported by Hui \& Lin~\cite{Hui:2024ggb}, who used a Cartesian-grid-based multigrid solver in the \et with tolerance $||\Theta||_\infty = 2.3\times10^{-8}$. Our effective tolerance of $1\times10^{-8}$ (i.e., $(3\times10^{-8})/3$) is about 2.3 times stricter.

Although $R_{\rm crit}$ depends on the convergence tolerance---tighter criteria yield smaller values due to the monotonic decrease of $\Theta(r,\theta,\phi)$ near the MOTS---the agreement across AH finders is excellent. These results demonstrate that, when supplied with high-resolution data and operated at strict tolerances, \bah achieves agreement with established AH finders to at least 8 significant digits, even in this demanding test case.

\section{Conclusions and Future Work}
\label{sec:conclusion}

We have introduced \bah, the \bhah Apparent Horizon Algorithm, the first open-source, infrastructure-agnostic library for AH finding in NR. \bah reformulates the elliptic MOTS equation as a damped nonlinear scalar wave equation on a 2-sphere, using a reference-metric formulation to manage coordinate singularities. It is the first method to use hyperbolic-relaxation flow for AH finding, and inherits the robustness of flow methods to poor initial guesses.

While hyperbolic relaxation is robust, it is typically slower than elliptic solvers. \bah mitigates this with two key strategies: a multigrid-like refinement that seeds fine-grid solves from coarse-grid solutions, and an over-relaxation technique adapted from numerical linear algebra. In addition, \bah is \openmp parallelized to scale across modern multicore CPUs.

These enhancements were critical for performance. In a challenging $q=4$ common horizon test case (Sec.~\ref{subsubsec:resultsbah_accelerations}, Fig.~\ref{fig:q4_3d}), they reduced total grid-point evaluations by over 90\% and achieved up to 26x speedups on eight cores over a na\"ive single-core implementation. Further, \bah outperformed the serial \ahfd by 21\% on 8 cores and 9.5\% on 16 cores, though it remained about twice as slow on a single core (Table~\ref{tab:timing_results}).

When tracking horizons in dynamical spacetimes, the accuracy of AH finders is typically limited by interpolation errors when transferring metric data from the host NR grid. For the GW150914-like BBH inspiral of Sec.~\ref{sec:gw150914}, these errors lead to relative horizon area discrepancies of ${\approx}\, 1.34 \times 10^{-5}$ as observed with \ahfd (Fig.~\ref{fig:bbh_area_comparison}), consistent with the accuracy reported in the abstract of the \ahfd announcement paper~\cite{Thornburg:2003sf}. With similar tolerances, \bah---using extrapolated initial guesses and optimized interpolation---runs ${\sim}\,$2.1x faster than \ahfd. At its default tolerance $m_{\rm scale} ||\Theta||_\infty=10^{-5}$, \bah yields relative area errors of ${\approx}\, 4.58 \times 10^{-6}$ and horizon trajectories agreeing with \ahfd to within $\Delta r/M {\sim}\, 10^{-6}$ (Fig.~\ref{fig:bbh_trajectory}), showing that \bah's speed advantage does not compromise accuracy.

A key strength of \bah is its invariance to overall system scale: all dimensionful tolerances and parameters are set relative to each horizon mass. This allows \bah to reliably compute normalized horizon areas across 16 orders of magnitude in mass---a regime where the tested \ahfd implementation fails or deviates significantly (Sec.~\ref{subsubsec:results_scale_invariance}, Table~\ref{tab:mass_area}).

Precision tests of the three-throat Brill--Lindquist configuration (Sec.~\ref{subsec:three_bh_crit}) show that \bah, when run with high internal resolution and strict tolerances, reproduces the value of $R_{\rm crit}$ obtained by other state-of-the-art AH finders to all significant digits.

Being an infrastructure-agnostic library, \bah has been successfully integrated into both the \et and \bhah NR frameworks, demonstrating its adaptability. These results further establish hyperbolic relaxation as a fast, accurate, and scale-invariant method for AH finding---combining the robustness of flow-based approaches with performance that matches or exceeds that of \ahfd.

Looking ahead, several developments will enhance \bah's capabilities. Planned improvements to the \bhah and \et implementations include spatial masking, allowing horizon interiors to be excised or GRHD/GRMHD fields quenched inside black holes. We aim to make the BBH mode more robust for automatic common horizon detection in both the \et and \bhah. Efforts are also underway to integrate \bah into additional NR infrastructures beyond the \et and \bhah.

Performance and algorithmic optimizations are also a priority. A major goal is to enable GPU support for \bah, building on recent success running \nell on CUDA-enabled GPUs~\cite{Tootle:2025ikk}. We also plan to explore new grid structures, such as tilted ellipsoidal coordinates, to better sample spacetime fields around spinning black holes. On the physics front, we plan to implement the isolated- and dynamical-horizon formalisms~\cite{Ashtekar:2002ag,Ashtekar:2003hk,Ashtekar:2004cn} to enable advanced quasi-local diagnostics (e.g., spin, mass, and energy or angular momentum fluxes).

The foundational role of hyperbolic relaxation in NR, established decades ago with the Gamma-driver shift conditions~\cite{Alcubierre:2000yz,Campanelli:2005dd}, has paved the way for its application in standalone initial-data solvers~\cite{Ruter:2017iph,Assumpcao:2021fhq,Tootle:2025ikk}. \bah's success in AH finding further demonstrates the power of this approach for solving elliptic PDEs in NR. Crucially, its multigrid-inspired and over-relaxation techniques are broadly applicable and poised to enhance performance across other current and potential hyperbolic relaxation applications, such as initial-data construction, constraint damping, and inverse-curl operations~\cite{Silberman:2018ioy}. We aim to continue developing \bah and explore the application of its acceleration techniques to other key problems in NR.

\section*{Acknowledgments}

We would like to thank W.~K.~Black, S.~Brandt, P.~Diener, M.~Fernando, R.~Haas, N.~Jadoo, B.~J.~Kelly, S.~T.~McWilliams, J.~Miller, S.~C.~Noble, E.~Schnetter, H.~Sundar, and several others for helpful discussions in the preparation of this manuscript. Z.B.E. gratefully acknowledges support from NSF awards PHY-2110352/2508377, PHY-2409654, OAC-2004311/2227105, OAC-2411068, and AST-2108072/2227080, as well as NASA awards ISFM-80NSSC18K0538, TCAN-80NSSC18K1488, and ATP-80NSSC22K1898. TA is thankful for support from NSF grants OAC-2229652 and AST-2108269. L.R.W. gratefully acknowledges support from NASA award LPS-80NSSC24K0360. Z.B.E. and ST received additional support from NASA award ATP-80NSSC22K1898 and the University of Idaho P3-R1 Initiative. This research made use of the resources of the High Performance Computing Center at Idaho National Laboratory, which is supported by the Office of Nuclear Energy of the U.S. Department of Energy and the Nuclear Science User Facilities under Contract No. DE-AC07-05ID14517. In addition, it made use of the Falcon~\cite{IdahoC3Plus3_2022_Falcon} supercomputer, operated by the Idaho C3+3 Collaboration.

\section*{References}

\bibliographystyle{iopart-num}

\bibliography{references}

\end{document}